\documentclass[fleqn,10pt,twocolumn]{ICCAS2017}

\usepackage{graphicx}
\usepackage{caption}
\usepackage{subcaption}
\usepackage{subfig}
\usepackage{amsfonts}
\usepackage{comment}
\usepackage{amssymb}
\usepackage{amsmath}
\usepackage{mdwlist}
\usepackage{rotating}
\usepackage{color}
\usepackage{soul}
\usepackage{cite}

\begin{document}

\title{Adaptive Regulation of Sampling Rates for Power Efficient \\ Embedded 
Control System Design}

% %%

% \author{}
% \affils{}

\author{Rajorshee Raha \vspace{-0.2cm} }
\affils{Department of Computer Science and Engineering, Indian Institute of 
Technology Kharagpur, INDIA \\
rajorshee.raha@cse.iitkgp.ernet.in
\vspace{-0.2cm}
}
\abstract{ 
In recent times adaptive regulation of sampling rates has gained significant 
attention in research community and researchers has demonstrated it's 
effectiveness in embedded control applications from different perspectives. In 
low power embedded control systems, the sampling rate of the control tasks has 
a direct relationship with control performance and power consumption. In this 
paper, we investigate the possibility of improving the power efficiency of low 
power embedded control systems by regulating the sampling rate of the control 
tasks. On this regard, we present algorithmic approaches for on-line regulation 
of sampling rates of the control task under some power-performance trade-off. 
We present elaborative results which demonstrate the efficacy of our proposed 
approaches.
\vspace{-0.3cm}}


\maketitle

%% 
% \vspace{-0.4cm}
%%%%%%%%%%%%
\section{Introduction}
Today majority of the embedded control applications are implemented using some 
digital platform, where the control tasks generally rely on periodic sampling 
and control. Traditionally the design of the control tasks are done using fixed 
and pre-defined sampling rate for the controller \cite{discretecontrolbook} and 
these sampling rate's are chosen in such a way, that the controller can 
guarantee the desired level of control performance {\em under all operating 
conditions}. In recent times adaptive regulation of sampling rates of the 
control task \cite{cervin2011paper, rr_msc15, rr_iccsce14, rr_dac14, 
rr_icstcc16, chinease, Cooperative, 2016_tc} has gained significant attention in 
research community, where researchers has demonstrated the trade-off between 
computational resources and control performance using adaptive sampling in 
different contexts.
 
Moreover with the increasing popularity of low power cyber-physical systems, 
researchers have looked into various types of power related issues, specially in 
the field of wearable technology and wireless network control systems 
\cite{wireless_control_survey, wireless_control_survey_1, vmnet}. In many low 
power embedded control system applications, power efficiency is one of the major 
area of concern \cite{power_wsn, power_sampling, power_sampling_1, 
recent_trends_power_2} as the sensors, actuators and electronic control units 
(ECU) can be battery powered with either individual or shared power supplies.

In \cite{cervin2011paper, rr_msc15} researchers had demonstrated that one can 
achieve better utilization of shared computational resources by adaptively 
regulating the sampling rates in response to the disturbance level at the 
plant. In software based control, a higher sampling rate indicates a higher 
rate of sensing and actuation of the control tasks and thus have direct effect 
in power consumption. Thus, in scenarios where the system operates under 
high disturbance levels, a rise in sampling rate in response to a disturbance 
might affect the energy efficiency of the system and, in situations where the 
disturbances are rare, a slight rise in sampling rate in response to a 
disturbance does not have any significant affect on the systems power 
efficiency.

In a recent work \cite{rr_esl16}, we had used this philosophy to study the 
trade-off between control performance and power under adaptive regulation of the 
sampling rate. The work presented in \cite{rr_esl16} considers the disturbance 
patterns generated from legacy data as an input to the problem and the control 
strategy adapts with different energy budget or energy efficiency levels. In 
\cite{rr_esl16}, we considered the disturbance patterns in different forms and 
proposed a methodology for choosing the best possible multi-rate controller 
which guarantees optimal control performance under certain energy budget.

Although the methodology presented in \cite{rr_esl16} provides an optimal 
solution, but is computationally very expensive because of the underling 
exhaustive search technique, which makes it practically hard to realize 
specially in cases where the computations are to be performed on-the-fly. In 
order to make the methodology computationally more efficient, some algorithmic 
improvement needs to be crafted such that it becomes feasible for cases where 
the computations are to be performed on-the-fly. In this paper, we extend the 
idea of the work presented in \cite{rr_esl16} and propose the following enabling 
contributions:
\begin{itemize}
  \item We propose a framework for on-line adaptive regulation of sampling 
rates under power-performance trade-off.

  \item We propose algorithmic approaches to achieve our goal of efficient 
on-line adaptive regulation of sampling rates, under such power performance 
trade-off.
\end{itemize}
Further, we present elaborative results which demonstrate the efficiency of our 
proposed approach. Next, we outline the relevant background details.
%%
% \vspace{-0.4cm}
\section{Background Study} \label{background}
In this paper, we consider the continuous-time linear systems (see 
\cite{discretecontrolbook, cervin2011paper, rr_msc15}) which are described as: 
\begin{equation} 
\label{Eq:plant}
\begin{array}{c}
\dot{x}(t) = A x(t) + B u(t) + v_{c}(t) \\
y(t) = C x(t) + D u(t) + e(t) \\
\end{array}
\end{equation}
here, $x$ is the plant state, $u$ is the input from the controller, $y$ is 
the output of the plant. $A,~ B,~ C,~ D$ are the system matrices. $v_{c}$ is a 
continuous time white noise process having zero mean and covariance matrix 
$R_{1c}$ and $e$ is a discrete-time Gaussian white noise process with zero mean 
and variance $R_{2}$. The covariance matrix $R_{1c}$, can be written as, 
$R_{1c}=r(t).R_c$. Here, $R_c$ is the nominal noise co-variance matrix and 
$r(t)$ is a time varying non negative scalar. The control performance of the 
control loops for such plants operating under the influence of Gaussian 
noise,are generally measured by a quadratic cost function given as (see 
\cite{cervin2011paper, discretecontrolbook, chinease, Cooperative}),
% 
% J=E{ lim τ→∞ 1 τ  τ 0 [x T ,u T]Qxu [xu]dt}
\begin{equation}\label{Eq:cost}
\begin{array}{c}
J= E \{ \underset{t \rightarrow \infty}{\lim} \frac{1}{t} \int^t_0 [ x^T, u^T ] 
Q_{xu} \begin{bmatrix} x \\ u \end{bmatrix} dt \}
\end{array}
\end{equation}
here $Q_{xu}$'s is a positive semidefinite matrix. The discrete-time 
representation of the quadratic control cost function is discussed in 
\cite{discretecontrolbook, cervin2011paper, rr_msc15, chinease, Cooperative, 
2016_tc}. The nature of the quadratic cost function is such that, a lower value 
of this cost function signifies a better level of control performance.

A linear quadratic Gaussian (LQG) controller \cite{discretecontrolbook} is 
optimal for this type of cost function. In optimal control theory, the LQG 
control problem is concerned with uncertain linear systems which are disturbed 
by additive white Gaussian noise and the control problem is optimized subject to 
such quadratic costs. The LQG control is very popular in control theoretic 
domain because of it's unique solution and moreover the feedback control law is 
easy to implement.

Generally, the sampling rate of any discrete time control system signifies a 
sense-compute-actuate cycle. Further, the periodicity of the software based 
control tasks are chosen from a finite set of sampling rates 
realizable/supported by the embedded computing platform. Let such a finite set 
of sampling rates is given as, 
$H = \{h_1, ..., h_n\}, h_1< \dots <h_n$. 
Let an execution pattern of a controller which uses multi-rate sampling 
strategy over a period of time $\tau$, is denoted as $\alpha$ and for such a 
execution pattern, let $T_1, \ldots, T_m$ be the maximal intervals within $\tau$ 
where the sampling rate is uniform. Therefore $\tau$ can be expressed as, 
$\tau=\sum_{j=1}^{m}T_j$, where $h_j$ denote the sampling rate used in the 
interval $T_j$. Further, the control performance cost of the overall system for 
such an execution pattern $\alpha$ is over the time window $\tau$ is given as, 
\vspace{-0.2cm}
\[{\mathcal{J}} = \frac{\sum_{j=1}^{m}  J(T_j, h_j) \times T_j}{\tau}\] 
where, $J(T_j, h_j)$ represents the control performance cost of a closed 
loop system with sampling rate $h_j$ deployed for a time interval of $T_j$. 
Moreover, the total energy consumption for the control execution pattern 
$\alpha$ over the time window $\tau$ is given as,
\vspace{-0.2cm}
\[E = \sum_{j=1}^{m} \lfloor \frac{T_j}{h_j}\rfloor \times\phi\] 
where $\phi$ is the component of energy consumed per sample in a 
control (sense-compute-actuate) cycle. Practically, the estimation of 
$\phi$ can be done using various techniques \cite{vmnet, Panigrahi} in 
different contexts. 

% \subsection{Brief Outline: Exhaustive Search Approach \cite{rr_esl16}}
In \cite{rr_esl16}, we proposed a methodology for synthesizing multi-rate 
controllers which switches between a pre-defined set of sampling rates, 
considering the disturbance patterns in different forms as an input based on 
some disturbance levels and the overall systems is constrained under some given 
energy budget. The disturbance levels were discretized as a finite set, 
$L=\{1,2, \ldots, k\}$. 

Apart from the definition of the controller and the knowledge about the 
disturbance pattern the system has the information about an energy budget given 
as $\langle {\cal E}, {\cal T} \rangle$, which specifies that the system can 
spent at most ${\cal E}$ units of energy for a duration of next ${\cal T}$ units 
of time. Estimate of battery life for determining the energy budget, can be 
calculated using existing well known methodologies \cite{Panigrahi}. In order to 
find the multi-rate controllers, a base line exhaustive search approach was 
proposed in \cite{rr_esl16}, which in turn is used to find a solution for the 
different scenarios as per the knowledge about the disturbance levels. 

A multi-rate controller chooses a sampling rate for each disturbance 
level, thus the total possible choice of multi-rate controllers are given as 
$|H|^{|L|}$. Given the knowledge about the disturbance pattern and an energy 
budget $\langle {\cal E}, {\cal T} \rangle$, the exhaustive/brute-force search 
approach finds the total control cost and the total energy consumption over the 
time  window ${\cal T}$ for all $|H|^{|L|}$ possible choice of multi-rate 
controllers and thereafter chooses the multi-rate option which promises the best 
control cost within the energy budget.

As the total possible choice of multi-rate options is given as $|H|^{|L|}$, thus 
with the increase in the elements in the sets $H$ and $L$, the total possible 
choice of multi-rate option and the computational overheads are expected to grow 
exponentially and therefore makes the approach \cite{rr_esl16} practically hard 
to realize or nearly infeasible specially in cases where the computations are to 
be performed on-the-fly.

\section{Methodology Outline} \label{methodology}
Our goal is to propose a computationally efficient power-aware strategy for 
regulating the sampling rates of the control tasks on-the fly in response to 
different level of criticality. The proposed sampling rate regulation framework 
is outlined as,
\begin{itemize}

 \item We consider that, initially the software based controller's does not 
have any pre-specified disturbance pattern as an input.

  \item Therefore initially the system starts with the most frequent choice of 
sampling rate, from the given choice of sampling rate, $H$ and further decides 
it's future actions by learning from the system's behavior. 

  \item  In order to do so, it maintains a recent history of the disturbances 
encountered over a finite time window, $T_h$.

  \item For the sampling rate regulation, we propose some algorithmic 
approaches for finding the appropriate multi-rate controller, ${\cal M}: L 
\rightarrow H$, here, the function specifies a sampling rate, $h \in H$, for 
each disturbance level, $l \in L$. 

  \item The regulation of the sampling rate of the control tasks are decided 
using our proposed algorithmic mechanism (Approach-I and Approach-II), which is 
the heart of our work and are discussed in Section \ref{dominance}~~~  and 
\ref{greedy}~~~  in detail.

  \item Thus, given the knowledge about the disturbance pattern derived based on 
the past history of events and depending upon the energy budget or energy 
efficiency levels, our proposed methodology (Approach-I/Approach-II) regulates 
the sampling rate of the control tasks on-the-fly while achieving desired level 
of control performance.

\end{itemize}

\subsection{Off-line Computations}\label{offline}
To reduce the computational overhead during the on-line computations, we 
pre-compute the control cost and the power consumption for different choice of 
sampling rates, $h_i \in H$ and different disturbance levels given as $l_j 
\in L$ ($1\leq i\leq n, 1\leq j\leq k$). These values are calculated and stored 
in some tabular form, where a table, ${\cal C}T[H][L]$ stores the control costs 
(computed using the quadratic cost function) for each sampling rate $h_i \in 
H$ and for each of the disturbance levels given as $l_j \in L$, and a table, 
${\cal P}T[H]$ stores the power consumption with respect to each sampling 
rate $h_i \in H$. 

% \vspace{-0.5cm}
\subsection{Approach-I}\label{dominance}
For the set of sampling rates $H=\{h_1, \dots h_n\}$ and disturbance levels 
$L = \{1, ..., k\}$, each choice of multi-rate controller out of the all 
possible choice given as $|H|^{|L|}$, can be characterized by the vector 
$\langle {\cal M} (1), ... ,{\cal M}(k) \rangle$ which maps the disturbance 
levels to a choice of sampling rates. Let ${\cal S}$ represent the set of all 
possible choice of multi-rate controllers given as $|H|^{|L|}$. 

We propose an intelligent pruning technique to reduce the search space. Using 
the notion of dominance, we introduce following pruning technique:
\begin{itemize}

  \item If a choice of multi-rate controller $\langle {\cal M} (1),\dots,{\cal 
M}(k)\rangle$ exceeds the energy budget, then we can prune all options of the 
form $\langle {\cal M}'(1),\dots,{\cal M}'(k) \rangle$ such that ${\cal M}' (j) 
\leq {\cal M} (j)$, for all $1 \leq j \leq k$, because all these options will 
also exceed the energy budget.

  \item  If a choice of multi-rate controller $\langle {\cal M} (1), ... ,{\cal 
M}(k) \rangle$ satisfies the energy budget, then we can prune all distinct 
options of the form $\langle {\cal M}' (1), ... ,{\cal M}'(k) \rangle$ such that 
${\cal M}(j) \leq {\cal M}'(j)$, $1 \leq j \leq k$, because none of these 
options will improve the control performance over $\langle {\cal M} (1), ... 
,{\cal M}(k) \rangle$.

\end{itemize}

Therefore, given the knowledge about the disturbance pattern derived based on 
the past history of events and an energy budget $\langle {\cal E}, {\cal T} 
\rangle$, initially this approach computes the total control cost and the total 
energy requirement (using the precomputed datasets ${\cal C} T[H][L]$ and ${\cal 
P} T[H]$ as discussed in Section \ref{offline}~~~) for the different choice of 
multi-rate options. Further using the above mentioned pruning technique, it 
prunes out the irrelevant choices of multi-rate controllers to get a filtered 
set of multi-rate options ${\cal S}'$, such that ${\cal S}' \subset {\cal S}$ 
and $|{\cal S}'| < |{\cal S}|$. 

Next, from this set of multi-rate options, ${\cal S}'$, we choose the switching 
function which promises the best control performance (lowest control cost) and 
also does not exceed the energy budget. This selected multi-rate controller is 
the one that optimizes control performance under the desired energy budget. In 
Section \ref{results}~, we provide illustrative results which demonstrate the 
benefit of our proposed Approach-I in terms of computational efficiency and 
control performance.
\subsection{Approach-II} \label{greedy}
Although, our proposed Approach-I reduces the search space to some extent and 
promises better computational efficiency compared to the exhaustive/brute-force 
search methodology presented in \cite{rr_esl16}, still the reduced search space 
can be somewhat large and therefore may still be computationally expensive for 
larger data sets. Hence, in order to further optimize the computational 
efficiency, we propose another approach (inspired from fractional knapsack 
problem), the steps of which are discussed next.

Given an energy budget $\langle {\cal E}, {\cal T} \rangle$ and knowledge about 
the disturbance pattern derived based on the past history of events, initially 
it computes the total control cost, ${\cal CC}Total[h_i][l_j]$, and the total 
energy consumption, ${\cal EC} Total[h_i]$, over the finite time interval ${\cal 
T}$, with the help of the pre-computed tables ${\cal C} T[H][L]$ and ${\cal P} 
T[H]$ as discussed in Section \ref{offline}~~. Next, we define a function, 
${\cal B}$ computed as ${\cal B} = 1/({\cal CC}Total[h_i][l_j] \times {\cal 
EC}Total[h_i])$, which defines the profit of using a particular sampling mode in 
terms of control performance (maximum performance) and energy (minimum power 
requirement). As, the nature of the quadratic cost function (Section 
\ref{background}~ ) is such that, a lower value of this cost function indicates 
a better level of control performance, hence we take a inverse of the control 
cost to represent maximum profit.

All these values are stored in some tabular form, where each table ${\cal 
B}TAB[H][l_j]$, will correspond to a specific disturbance levels, $l_j\in L$ 
and store the control cost, the energy cost, and the profit function value, 
${\cal B}$ corresponding to each choice of sampling rate $h_i \in H$, $1\leq i 
\leq n$, as shown in Table \ref{tab:knapsack}.
\begin{table}[h]
\vspace{-0.2cm}
\caption{${\cal B}TAB[H][l_j]$}
\vspace{-0.3cm}
\centering
% \resizebox{\linewidth}{!}{
\resizebox{\linewidth}{0.025\textwidth}{
\begin{tabular}{|c|c|c|c|} \hline
Sampling Periods & Control Cost & Energy Cost & Profit Function, (${\cal B}$) \\
%  & Cost & Cost &  \\
 \hline
$h_i \in H$ & ${\cal CC}Total[h_i][l_j]$ & ${\cal EC}Total[h_i]$ & 
$1/({\cal CC}Total[h_i][l_j] \times {\cal EC}Total[h_i])$ \\ \hline
\end{tabular}
}
\vspace{-0.35cm}
\label{tab:knapsack}
\end{table}

A multi-rate controller, ${\cal M}: L \rightarrow H$, needs to choose a sampling 
rate for each disturbance level. As, the energy budget is a constraint on the 
energy consumption, thus we try to choose the multi-rate controller which 
guarantees maximum collective profit within the given energy budget. Therefore, 
we consider the elements of each of these tables, ${\cal B}TAB[H][l_j]$, and 
sort those elements in descending order in terms of the profit function, ${\cal 
B}$, and further store the rearranged elements associated with each of the 
profit function (e.g. sampling rate, total control cost and total energy cost) 
in tabular format in table $STAB[H][l_j]$. Thus, we have $k$ (=$|L|$), such 
sorted tables in terms of profit function for each of the disturbance levels and 
each of them will have $n$ (=$|H|$), entries corresponding to sampling rates. 
% %%
% \begin{table}[h]
% % \vspace{-0.3cm}
% \caption{$STAB [j] [i]$: Sorted Elements - For Each disturbance Levels}
% % \vspace{-0.2cm}
% \centering
% % \resizebox{\linewidth}{!}{
% \begin{tabular}{|c|c|c|} \hline
% Sampling Periods & Control Cost & Energy Cost \\ \hline
% $h'_i$ & $CCTotal[i][j]$ & $ECTotal[i]$ \\ \hline
% \end{tabular}
% % }
% % \vspace{-0.4cm}
% \label{tab:sort}
% \end{table}
% %%
Next, in order to find the appropriate choice of multi-rate controller, we 
search the possible combinations of multi-rate options in the following manner 
as given in Table \ref{tab:binary}.  

\begin{table}[h]
\vspace{-0.1cm}
\caption{Selection of appropriate multi-rate controller following binary search 
pattern}
\vspace{-0.2cm}
\centering
\resizebox{\linewidth}{!}{
\begin{tabular}{|c||c|c|c|c|c|} \hline
Choice & Mode-1 & Mode-2 & \dots & Mode-(k-1) & Mode-k\\ \hline
choice 1 & $STAB[1] [1]$ & $STAB[1] [2]$ & \dots & $STAB[1] [k-1]$ & $STAB[1] 
[k]$\\ \hline
choice 2 & $STAB[2] [1]$ & $STAB[1] [2]$ & \dots & $STAB[1] [k-1]$ & $STAB[1] 
[k]$\\ \hline
choice 3 & $STAB[1] [1]$ & $STAB[2] [2]$ & \dots & $STAB[1] [k-1]$ & $STAB[1] 
[k]$\\
\vdots& \vdots & \vdots & \vdots & \vdots & \vdots \\
\vdots & $STAB[1] [1]$ & $STAB[1] [2]$ & \dots & $STAB[1] [k-1]$ & $STAB[2] 
[k]$\\ 
\cline{2-6}
. & $STAB[2] [1]$ & $STAB[2] [2]$ & \dots & $STAB[2] [k-1]$ & $STAB[2] [k]$\\
\vdots & \vdots & \vdots & \vdots & \vdots & \vdots \\
\vdots & $STAB[n] [1]$ & $STAB[n] [2]$ & \dots & $STAB[n] [k-1]$ & $STAB[n] 
[k]$\\ \hline
\end{tabular}
}
\vspace{-0.2cm}
\label{tab:binary}
\end{table}

Table \ref{tab:binary}, depicts a binary table representation, where initially 
the choice of multi-rate option with maximum collective profit is selected and 
thereafter binary tabular search pattern is followed. The benefit of following 
such a binary tabular search pattern is that, for the next choice, the sampling 
rate of only one mode has to be varied and the succeeding choice always 
guarantees next possible maximum profit. Following this search technique, we 
finally select the multi-rate controller that satisfies the energy budget in 
first place. In Section \ref{results}~, we present appropriate results which 
highlights the computational efficiency of this approach.
%%
%% Runtime Comparision Table
%%%%%%%%%%%%%%%%%%%%%%%%%%%%%%%%%%%%%%%%%%%%%%%%
\begin{table*}[!htb]
\vspace{-0.2cm}
\caption{Comparative Study: Computational Efficiency}
% \caption{Comparative Study of Computational Efficiency: Exhaustive approach 
% \cite{rr_esl16} v/s Proposed Approach I and II}
\vspace{-0.3cm}
\centering
% \resizebox{<width>}{<height>}
\resizebox{\linewidth}{0.11\textwidth}{
\begin{tabular}{|c|c|c|c|c|c|c|c|c|} \hline
Total choice of & \multicolumn{3}{c|}{Number of Sampling Combination Explored} 
& \multicolumn{3}{c|}{Run Time (in second)} & \multicolumn{2}{c|}{Run Time 
Ratio} \\ \cline {2-7} 

Sampling rates & Exhaustive & \multicolumn{2}{c|}{Proposed} & Exhaustive & 
\multicolumn{2}{c|}{Proposed} & \multicolumn{2}{c|}{w.r.t. Approach-II} \\ 
\cline {3-4} \cline {6-7} \cline{8-9}

$|H|$ & Search, \cite{rr_esl16} & Approach-I &  Approach-II & Search, 
\cite{rr_esl16} & Approach-I & Approach-II & Exhaustive, \cite{rr_esl16} & 
Approach-I \\ \hline

9   & 729          & 438          & 14   & 0.122849      & 0.073819     & 
0.002349 & 52.2984   & 31.4257 \\ \hline

16  & 4,096        & 2436         & 29   & 0.184911      & 0.110068      & 
0.003220 & 57.42577  & 34.1826 \\ \hline

32  & 32,768       & 20,894       & 53   & 0.478642      & 0.305212      & 
0.004757 & 100.6184  & 64.1606 \\ \hline

80  & 5,12,000     & 3,38,486     & 129  & 5.181520      & 3.425517      & 
0.005768 & 898.3217  & 593.8829 \\ \hline

160 & 40,96,000    & 27,39,574    & 257  & 40.883443     & 27.344521     & 
0.007158 & 5711.5734 & 3820.1342 \\ \hline

800 & 51,20,00,000 & 30,76,41,532 & 1281 & Out of Memory & Out of Memory & 
0.031091 &    -      & -  \\ \hline
\end{tabular}
}
\label{tab:runtime}
\vspace{-0.2cm}
\end{table*}
%%%%%%%%%%%%%%%%%%%%%%%%%%%%%%%%%%%%%%%%%%%%%%%%
\section{Results and Discussion}\label{results}
In this section, we present experimental results in support of out proposed 
methodology. The experiments were carried out using MATLAB, TrueTime simulator 
\cite{cervin_tool} and Jitterbug Toolbox \cite{jitterbug, using_jitterbug}. We 
considered a test case from the TrueTime distribution, consisting of a DC 
servo motor controlled through a wireless sensor network. The linear plant 
dynamics is given by,
\begin{equation}
\begin{array}{l}
\dot{x}(t) =
\begin{bmatrix} -1 & 0\\
		1 & 0
\end{bmatrix}
x(t) + 
\begin{bmatrix} 1 \\
		0 
\end{bmatrix}
 u(t) + v_{c}(t)\\
y(t) = \begin{bmatrix} 0 & 1000\\ \end{bmatrix} x(t) + 
      \begin{bmatrix} 0 \end{bmatrix} u(t) + e(t) \\
\end{array}
\end{equation}
Further, we assume that the embedded platform admits the following sampling 
rates, $H = \{10, 15, 20, \dots 90\}$ in milliseconds. The noise level is 
determined from the estimated (using residual variance estimator (RVE) 
\cite{cervin2011paper}) value, $\hat{r}(t)$ as follows, 
$L=1 : ~0 < \hat{r}(t) \leq 10$, $L=2: ~10 < \hat{r}(t) \leq 50$, $L=3: ~50 < 
\hat{r}(t) \leq 100$. 
The noise levels for $L = 1, 2, 3$ are referred to as {\em low}, {\em medium} 
and {\em high} respectively.

Initially, we compute ${\cal C}T[i][j],~ 1\leq i\leq n,~ 1\leq j\leq k$, 
following the steps of our methodology as described in Section 
\ref{offline}~. The calculations were done using Jitterbug Toolbox 
\cite{jitterbug, using_jitterbug} in MATLAB. 
Next, the entries of ${\cal P}T[i], 1\leq i\leq n$, were computed using the 
following steps. In this example, the peak power consumption is considered as 
{\em 100} $mW$, which corresponds to the shortest sampling rate, $h_1 = 10ms$, 
thus we have ${\cal P}T[1] = 100$. Next, we compute the remaining entries of the 
table as, ${\cal P}T[i] = (100*10)/h_i, \quad  h_i \in H, 1 < i\leq n$.

We consider a scheduling hyper-period $T_h$ as $100sec$. As discussed in 
Section \ref{methodology}~ , our proposed methodology initially selects the 
most frequent sampling rate ($10ms \in H$) for the first time window, $T_h$ and 
thereafter regulates the sampling rate by learning from the system's behavior. 
The knowledge about the disturbance pattern is derived based on the past history 
of events occurred during the past a scheduling hyper-period $T_h$. Next, given 
the desired energy efficiency levels and knowledge of the disturbance pattern, 
our proposed methodology computes the multi-rate controllers (using 
Approach-I/Approach-II as described in Section \ref{methodology}~~) for the next 
time window, $T_h$ and thereafter follows the same steps.

\begin{figure}[h]
\vspace{-0.1cm}
\centering
\resizebox{\linewidth}{!}{
\includegraphics{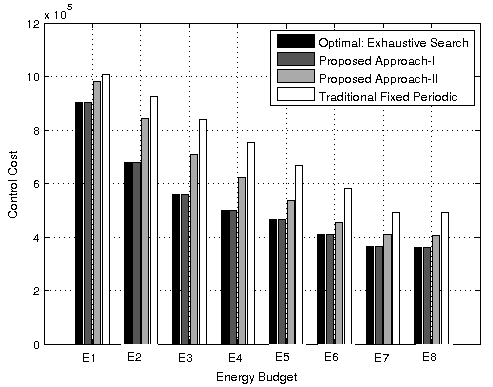}
}
\vspace{-0.6cm}
  \caption{Comparative Study: Control Cost}
  \vspace{-0.4cm}
\label{fig:c2}
\end{figure}

Next, we provide elaborative results while considering one such scheduling 
hyper-period, $T_h$, where the estimated noise level distribution is considered 
as $\langle low$=$70\%, medium$=$10\%, high$=$20\% \rangle$. Assuming such 
operating scenario, the multi-rate controllers are synthesized using our 
proposed Approach-I (Section \ref{dominance}~~) and Approach-II (Section 
\ref{greedy}~~ ). The multi-rate controllers were also synthesized using the 
brute-force approach presented in \cite{rr_esl16}. In Figure \ref{fig:c2}, 
we present a comparative study between controllers using traditional fixed 
sampling strategy ($h$=$50ms$) and controllers using multi-rate sampling 
strategy for different energy budget options.

It is evident from the results presented in Figure \ref{fig:c2}, that multi-rate 
controllers promises better control performance (a lower value of control cost 
indicates a better level of control performance) compared to the fixed sampling 
strategy (Note that the y-axis has a scale factor of $10^5$). The multi-rate 
controller's synthesized using our proposed Approach-I and the approach 
presented in \cite{rr_esl16} promises best control performance compared to the 
others. Whereas, the multi-rate controllers synthesized using Approach-II 
promises satisfactory level of control performance compared to the other 
approaches. Although our proposed Approach-II lags in terms of control 
performance, when compared with our proposed Approach-I and the exhaustive 
search approach presented in \cite{rr_esl16}, but it is advantageous over them 
in terms of computational efficiency which is showcased next.

We present runtime comparison for synthesizing the multi-rate controllers using 
our proposed approaches (Approach-I and Approach-II) and the approach presented 
in \cite{rr_esl16}. The experiments were carried out using MATLAB, in a system 
with Intel Core i7-2600 processor clocked at 3.40 GHz with 8 cores, with a 
primary memory of 8 GB (DDR3), with disc space of 400 GB and with Ubuntu 14.04 
LTS (64 bit) operating system. For demonstration purpose, the range of sampling 
rate is considered in between $H$=$\{10 \dots 90\} ms$ and the total number 
of noise levels is taken as $|L|$=$3$. The finding are highlighted in Table 
\ref{tab:runtime}. 

It may be observed from the results presented in Table \ref{tab:runtime} that, 
both of our proposed approaches (Approach-I and Approach-II) are advantageous in 
terms of computational efficiency (less number of sampling combination 
explored to find a solution and therefore shorter runtime) compared to the 
exhaustive search approach \cite{rr_esl16}. Further, it is evident from the 
results showcased in Table \ref{tab:runtime}, that our proposed Approach-II 
clearly has an edge over the exhaustive search approach \cite{rr_esl16} and our 
proposed Approach-I in terms of computational efficiency (significant runtime 
ratio, which is calculated by dividing the runtime achieved using other 
approach, with the runtime achieved using Approach-II, see Table 
\ref{tab:runtime}).

The runtime efficiency of our approach is very advantageous considering the 
configuration of the ECUs \cite{microgen, openecu} used in embedded 
applications. It may be concluded that, our proposed Approach-I promises optimal 
result in terms of control performance but is computationally slightly expensive 
than our proposed Approach-II, whereas our proposed Approach-II is 
computationally very efficient and promises satisfactory level of control 
performance. So clearly there is a trade-off between the two proposed approaches 
(Approach-I and Approach-II) in terms of computational efficiency and control 
performance and any one of them can be used effectively based on the embedded 
control application requirements and overall system configuration.

\begin{figure}[!htb]
\centering
 \begin{subfigure}[b]{\linewidth}
 % $\langle low$=$20\%, medium$=$10\%, high$=$70\%\rangle$
 \centering
  \includegraphics[height=1.6in, width=3.2in]{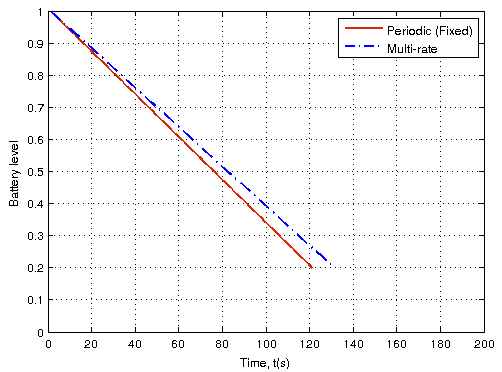}
     \vspace{-0.5cm}
  \caption{Scenario-I: where high noise level is encountered more often}
  \label{fig:c2_battery_1}
 \end{subfigure}
 \begin{subfigure}[b]{\linewidth}
%  $\langle low$=$70\%, medium$=$10\%, high$=$20\% \rangle$
 \centering
  \includegraphics[height=1.6in, width=3.2in]{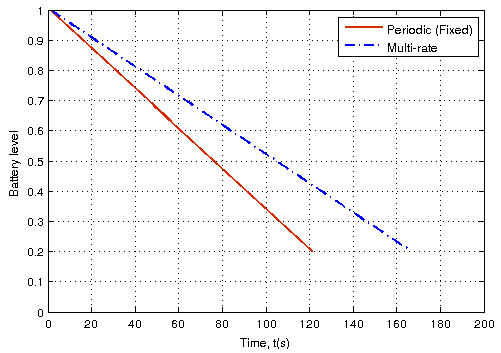}
     \vspace{-0.5cm}
  \caption{Scenario-II: where low noise level is encountered more often}
   \vspace{-0.3cm}
  \label{fig:c2_battery_2}
 \end{subfigure}
 \caption{Battery Life: Multi-rate v/s Fixed Periodic}
 \vspace{-0.1cm}
\end{figure}

Next, we showcase the power efficiency (possible extension in battery life) that 
can be achieved by using multi-rate controllers over fixed sampling strategy. We 
considered two different set of disturbance scenarios ``Scenario-I'' where high 
noise level is encountered more often and ``Scenario-II'' where low noise level 
is encountered more often. We employed our proposed methodology in these two 
different disturbance scenarios and calculated the power related benefits 
achieved compared to fixed sampling strategy and the findings are highlighted in 
Table \ref{tab:battery}. Further, the corresponding battery discharge results 
are shown in Figure \ref{fig:c2_battery_1} and \ref{fig:c2_battery_2}. 
\begin{table}[!htb]
\vspace{-0.1cm}
\caption{Comparative Study: Power Efficiency}
\vspace{-0.3cm}
\centering
\resizebox{\linewidth}{!}{
\begin{tabular}{|c|c|c|c|} \hline
Test & Reduction in Power & \multicolumn{2}{c|}{Battery Life 
Enhanced}\\ \cline{3-4}
Pattern & Requirements (in \%) & (in \%) & ($\approx$ in months) \\ \hline
I & $18.19$ & $7.82$ &  $4.71$ (nearly 5) \\ \hline % $4.71$
II & $50.18$ & $27.55$ & $16.65$ (nearly 17) \\ \hline % $16.53$
\end{tabular}
}
\vspace{-0.3cm}
\label{tab:battery}
\end{table}

Results presented in Table \ref{tab:battery}, Figure \ref{fig:c2_battery_1} and 
\ref{fig:c2_battery_2}, highlight that the multi-rate sampling strategy results 
in significant saving in power requirements and enhances the battery life time, 
therefore promising power efficiency.

\section{Conclusion}
This paper investigates the possibility of improving power efficiency of low 
power embedded control system by adaptive regulation of sampling rates of the 
control task.  We present computationally efficient approaches for on-line 
regulation of sampling rates under power performance trade-off. We report 
encouraging results in terms of control performance, power efficiency and 
computational efficiency. We believe that the framework proposed in this paper 
for adaptive regulation of sampling rates of the control task is novel in 
treatment of the problem and further motivates future research in these lines.

%%
% \vspace{-0.05cm}
\bibliographystyle{ieeetr}
\bibliography{paper_draft}

\begin{thebibliography}{10}

\bibitem{discretecontrolbook}
K.~{\AA}str{\"o}m and B.~Wittenmark, {\em Computer controlled systems: theory
  and design}.
\newblock Prentice-Hall, 1984.

\bibitem{cervin2011paper}
A.~Cervin, M.~Velasco, P.~Marti, and A.~Camacho, ``Optimal online sampling
  period assignment: Theory and experiments,'' {\em IEEE Transactions on
  Control Systems Technology,}, vol.~19, no.~4, pp.~902--910, 2011.

\bibitem{rr_msc15}
R.~Raha, S.~Dey, and P.~Dasgupta, ``Adaptive sharing of sampling rates among
  software based controllers,'' in {\em International Symposium on Intelligent
  Control (ISIC) under 2015 IEEE Multi Conference on System and Control (MSC)},
  (Sydney, Australia), pp.~688--694, Sep 2015.

\bibitem{rr_iccsce14}
R.~Raha, A.~Hazra, A.~Mondal, S.~Dey, P.~P. Chakrabarti, and P.~Dasgupta,
  ``Synthesis of sampling modes for adaptive control,'' in {\em 4th
  International Conference on Control System, Computing and Engineering
  (ICCSCE)}, (Penang, Malaysia), pp.~294--299, Nov 2014.

\bibitem{rr_dac14}
R.~Raha, S.~Dey, P.~P. Chakrabarti, and P.~Dasgupta, ``Multi-mode sampling
  period selection for embedded real time control[poster],'' in {\em 51st
  Design Automation Conference (DAC)}, (San Fransisco, CA), Jul 2014.

\bibitem{rr_icstcc16}
R.~Raha, ``Synthesis of sampling modes of multi-rate systems for guaranteed
  stability,'' in {\em 20th International Conference on System Theory, Control
  and Computing (ICSTCC),}, (Sinaia, Romania), pp.~398--403, Oct 2016.

\bibitem{chinease}
C.~Du, L.~Tan, and Y.~Dong, ``Period selection for integrated controller tasks
  in cyber-physical systems,'' {\em Chinese Journal of Aeronautics}, vol.~28,
  no.~3, pp.~894 -- 902, 2015.

\bibitem{Cooperative}
V.~Moraes, M.~Jungers, U.~Moreno, and E.~Castelan, ``Sampling period
  assignment: A cooperative design approach,'' in {\em 2014 IEEE 53rd Annual
  Conference on Decision and Control (CDC)}, pp.~4361--4366, Dec 2014.

\bibitem{2016_tc}
P.~Deng, Q.~Zhu, A.~Davare, A.~Mourikis, X.~Liu, and M.~D. Natale, ``An
  efficient control-driven period optimization algorithm for distributed
  real-time systems,'' {\em IEEE Transactions on Computers}, vol.~65,
  pp.~3552--3566, Dec 2016.

\bibitem{wireless_control_survey}
M.~Pajic, S.~Sundaram, G.~Pappas, and R.~Mangharam, ``The wireless control
  network: A new approach for control over networks,'' {\em IEEE Transactions
  on Automatic Control,}, vol.~56, pp.~2305--2318, Oct 2011.

\bibitem{wireless_control_survey_1}
R.~Alur, A.~D'Innocenzo, K.~Johansson, G.~Pappas, and G.~Weiss, ``Modeling and
  analysis of multi-hop control networks,'' in {\em 15th IEEE Real-Time and
  Embedded Technology and Applications Symposium, RTAS 2009.}, pp.~223--232,
  April 2009.

\bibitem{vmnet}
H.~Wu, Q.~Luo, P.~Zheng, and L.~M. Ni, ``Vmnet: Realistic emulation of wireless
  sensor networks,'' {\em IEEE Trans. Parallel Distrib. Syst.}, vol.~18,
  pp.~277--288, Feb 2007.

\bibitem{power_wsn}
M.~Sichitiu, ``Cross-layer scheduling for power efficiency in wireless sensor
  networks,'' in {\em Twenty-third AnnualJoint Conference of the IEEE Computer
  and Communications Societies, INFOCOM 2004.}, vol.~3, pp.~1740--1750 vol.3,
  March 2004.

\bibitem{power_sampling}
H.~Wu and Q.~Luo, ``Supporting adaptive sampling in wireless sensor networks,''
  in {\em Wireless Communications and Networking Conference (WCNC)},
  pp.~3442--3447, March 2007.

\bibitem{power_sampling_1}
C.~Alippi, G.~Anastasi, C.~Galperti, F.~Mancini, and M.~Roveri, ``Adaptive
  sampling for energy conservation in wireless sensor networks for snow
  monitoring applications,'' in {\em IEEE International Conference on Mobile
  Adhoc and Sensor Systems ( MASS)}, pp.~1--6, Oct 2007.

\bibitem{recent_trends_power_2}
P.~Reiner and T.~Xie, ``Current trends in power aware design,'' in {\em IECON
  2012-38th Annual Conference on IEEE Industrial Electronics Society},
  pp.~6280--6284, IEEE, 2012.

\bibitem{rr_esl16}
R.~Raha, S.~Dutta, S.~Dey, and P.~Dasgupta, ``Multi-rate sampling for
  power-performance trade-off in embedded control,'' {\em IEEE Embedded Systems
  Letters (ESL)}, vol.~8, no.~4, pp.~77--80, 2016.

\bibitem{Panigrahi}
D.~Panigrahi, C.~Chiasserini, S.~Dey, R.~Rao, A.~Raghunathan, and K.~Lahiri,
  ``Battery life estimation of mobile embedded systems,'' in {\em 14th
  International Conference on VLSI Design}, pp.~57--63, 2001.

\bibitem{cervin_tool}
A.~Cervin, D.~Henriksson, B.~Lincoln, J.~Eker, and K.~Arzen, ``How does control
  timing affect performance? analysis and simulation of timing using jitterbug
  and truetime,'' {\em IEEE Control Systems Magazine}, vol.~23, pp.~16--30,
  June 2003.

\bibitem{jitterbug}
B.~Lincoln and A.~Cervin, ``Jitterbug: {A} tool for analysis of real-time
  control performance,'' in {\em Proceedings of the 41st IEEE Conference on
  Decision and Control}, (Las Vegas, NV), Dec 2002.

\bibitem{using_jitterbug}
A.~Cervin, ``Using {Jitterbug} to derive control loop timing requirements,'' in
  {\em Proceedings of CERTS'03 -- Co-Design of Embedded Real-Time Systems
  Workshop}, (Porto, Portugal), Jul 2003.

\bibitem{microgen}
``Microgen, [online]:www.in.mathworks.com/products
  /connections/product\_detail/product\_35678.html.''

\bibitem{openecu}
``Pi innovo openecu, [online]:www.autonomoustuff.
  com/pi\-innovo\-openecu.html.''

\end{thebibliography}
% \vspace{-0.2cm}
\end{document}